     
\font\twelverm=cmr10 scaled 1200    \font\twelvei=cmmi10 scaled 1200
\font\twelvesy=cmsy10 scaled 1200   \font\twelveex=cmex10 scaled 1200
\font\twelvebf=cmbx10 scaled 1200   \font\twelvesl=cmsl10 scaled 1200
\font\twelvett=cmtt10 scaled 1200   \font\twelveit=cmti10 scaled 1200
\font\twelvesc=cmcsc10 scaled 1200  
\skewchar\twelvei='177   \skewchar\twelvesy='60
     
     
\def\twelvepoint{\normalbaselineskip=12.4pt plus 0.1pt minus 0.1pt
  \abovedisplayskip 12.4pt plus 3pt minus 9pt
  \belowdisplayskip 12.4pt plus 3pt minus 9pt
  \abovedisplayshortskip 0pt plus 3pt
  \belowdisplayshortskip 7.2pt plus 3pt minus 4pt
  \smallskipamount=3.6pt plus1.2pt minus1.2pt
  \medskipamount=7.2pt plus2.4pt minus2.4pt
  \bigskipamount=14.4pt plus4.8pt minus4.8pt
  \def\rm{\fam0\twelverm}          \def\it{\fam\itfam\twelveit}%
  \def\sl{\fam\slfam\twelvesl}     \def\bf{\fam\bffam\twelvebf}%
  \def\mit{\fam 1}                 \def\cal{\fam 2}%
  \def\sc{\twelvesc}               \def\tt{\twelvett}
  \def\sf{\twelvesf}
  \textfont0=\twelverm   \scriptfont0=\tenrm   \scriptscriptfont0=\sevenrm
  \textfont1=\twelvei    \scriptfont1=\teni    \scriptscriptfont1=\seveni
  \textfont2=\twelvesy   \scriptfont2=\tensy   \scriptscriptfont2=\sevensy
  \textfont3=\twelveex   \scriptfont3=\twelveex  \scriptscriptfont3=\twelveex
  \textfont\itfam=\twelveit
  \textfont\slfam=\twelvesl
  \textfont\bffam=\twelvebf \scriptfont\bffam=\tenbf
  \scriptscriptfont\bffam=\sevenbf
  \normalbaselines\rm}
     

     
\def\beginlinemode{\endmode
  \begingroup\parskip=0pt \obeylines\def\\{\par}\def\endmode{\par\endgroup}}
\def\beginparmode{\endmode
  \begingroup \def\endmode{\par\endgroup}}
\let\endmode=\par
{\obeylines\gdef\
{}}
\def\singlespace{\baselineskip=\normalbaselineskip}

\def\oneandahalfspace{\baselineskip=\normalbaselineskip
  \multiply\baselineskip by 3 \divide\baselineskip by 2}
\def\doublespace{\baselineskip=\normalbaselineskip \multiply\baselineskip by 2}

\newcount\firstpageno
\firstpageno=2
\footline={\ifnum\pageno<\firstpageno{\hfil}\else{\hfil\twelverm\folio\hfil}\fi}
\def\toppageno{\global\footline={\hfil}\global\headline
  ={\ifnum\pageno<\firstpageno{\hfil}\else{\hfil\twelverm\folio\hfil}\fi}}
\let\rawfootnote=\footnote              
\def\footnote#1#2{{\rm\singlespace\parindent=0pt\parskip=0pt
  \rawfootnote{#1}{#2\hfill\vrule height 0pt depth 6pt width 0pt}}}
\def\raggedcenter{\leftskip=4em plus 12em \rightskip=\leftskip
  \parindent=0pt \parfillskip=0pt \spaceskip=.3333em \xspaceskip=.5em
  \pretolerance=9999 \tolerance=9999
  \hyphenpenalty=9999 \exhyphenpenalty=9999 }
\def\dateline{\rightline{\ifcase\month\or
  January\or February\or March\or April\or May\or June\or
  July\or August\or September\or October\or November\or December\fi
  \space\number\year}}
\def\received{\vskip 3pt plus 0.2fill
 \centerline{\sl (Received\space\ifcase\month\or
  January\or February\or March\or April\or May\or June\or
  July\or August\or September\or October\or November\or December\fi
  \qquad, \number\year)}}
     
     
\hsize=6.5truein
\vsize=8.5truein  
\parskip=\medskipamount
\def\\{\cr}
\twelvepoint            
\doublespace            
\overfullrule=0pt       

\def\title                      
  {\null\vskip 3pt plus 0.2fill
   \beginlinemode \doublespace \raggedcenter \bf}
     
\def\author                     
  {\vskip 3pt plus 0.2fill \beginlinemode
   \singlespace \raggedcenter\sc}
     
\def\affil                      
  {\vskip 3pt plus 0.1fill \beginlinemode
   \oneandahalfspace \raggedcenter \sl}
     
\def\abstract                   
  {\vskip 3pt plus 0.3fill \beginparmode
   \singlespace ABSTRACT: }
     
\def\endtopmatter               
  {\endpage                     
   \body}
     
\def\body                       
  {\beginparmode}               
     
\def\head#1{                    
  \goodbreak\vskip 0.5truein    
  {\immediate\write16{#1}
   \raggedcenter \uppercase{#1}\par}
   \nobreak\vskip 0.25truein\nobreak}
     
\def\subhead#1{                 
  \vskip 0.25truein             
  {\raggedcenter {#1} \par}
   \nobreak\vskip 0.25truein\nobreak}
     
\def\beginitems{
\par\medskip\bgroup\def\i##1 {\item{##1}}\def\ii##1 {\itemitem{##1}}
\leftskip=36pt\parskip=0pt}
\def\enditems{\par\egroup}
     
\def\beneathrel#1\under#2{\mathrel{\mathop{#2}\limits_{#1}}}
     
\def\refto#1{$^{#1}$}           
     
\def\references                 
  {\head{References}            
   \beginparmode
   \frenchspacing \parindent=0pt \leftskip=1truecm
   \parskip=8pt plus 3pt \everypar{\hangindent=\parindent}}

\gdef\refis#1{\item{#1.\ }}                     
     
\gdef\journal#1, #2, #3, 1#4#5#6{               
    {\sl #1~}{\bf #2}, #3 (1#4#5#6)}            

\gdef\refa#1, #2, #3, #4, 1#5#6#7.{\noindent#1, #2 {\bf #3}, #4 (1#5#6#7).\rm} 

\gdef\refb#1, #2, #3, #4, 1#5#6#7.{\noindent#1 (1#5#6#7), #2 {\bf #3}, #4.\rm} 

\def\pr{\journal Phys.Rev., }

\def\endreferences{\body}

\def\endpage                    
  {\vfill\eject}
     
\def\endpaper                   
  {\endmode\vfill\supereject}

\def\ref#1{Ref.~#1}                     
\def\Ref#1{Ref.~#1}                     
\def\[#1]{[\cite{#1}]}
\def\cite#1{{#1}}
\def\(#1){(\call{#1})}
\def\call#1{{#1}}
\def\taghead#1{}
\def\frac#1#2{{#1 \over #2}}
\def\half{{\frac 12}}

\def\12{{1\over2}}

\catcode`@=11
\newcount\r@fcount \r@fcount=0
\newcount\r@fcurr
\immediate\newwrite\reffile
\newif\ifr@ffile\r@ffilefalse
\def\w@rnwrite#1{\ifr@ffile\immediate\write\reffile{#1}\fi\message{#1}}

\def\writer@f#1>>{}
\def\referencefile{
  \r@ffiletrue\immediate\openout\reffile=\jobname.ref%
  \def\writer@f##1>>{\ifr@ffile\immediate\write\reffile%
    {\noexpand\refis{##1} = \csname r@fnum##1\endcsname = %
     \expandafter\expandafter\expandafter\strip@t\expandafter%
     \meaning\csname r@ftext\csname r@fnum##1\endcsname\endcsname}\fi}%
  \def\strip@t##1>>{}}

\def\citeall#1{\xdef#1##1{#1{\noexpand\cite{##1}}}}
\def\cite#1{\each@rg\citer@nge{#1}}	

\def\each@rg#1#2{{\let\thecsname=#1\expandafter\first@rg#2,\end,}}
\def\first@rg#1,{\thecsname{#1}\apply@rg}	
\def\apply@rg#1,{\ifx\end#1\let\next=\relax
\else,\thecsname{#1}\let\next=\apply@rg\fi\next}

\def\citer@nge#1{\citedor@nge#1-\end-}	
\def\citer@ngeat#1\end-{#1}
\def\citedor@nge#1-#2-{\ifx\end#2\r@featspace#1 
  \else\citel@@p{#1}{#2}\citer@ngeat\fi}	
\def\citel@@p#1#2{\ifnum#1>#2{\errmessage{Reference range #1-#2\space is bad.}%
    \errhelp{If you cite a series of references by the notation M-N, then M and
    N must be integers, and N must be greater than or equal to M.}}\else%
 {\count0=#1\count1=#2\advance\count1 by1\relax\expandafter\r@fcite\the\count0,
  \loop\advance\count0 by1\relax
    \ifnum\count0<\count1,\expandafter\r@fcite\the\count0,%
  \repeat}\fi}

\def\r@featspace#1#2 {\r@fcite#1#2,}	
\def\r@fcite#1,{\ifuncit@d{#1}
    \newr@f{#1}%
    \expandafter\gdef\csname r@ftext\number\r@fcount\endcsname%
                     {\message{Reference #1 to be supplied.}%
                      \writer@f#1>>#1 to be supplied.\par}%
 \fi%
 \csname r@fnum#1\endcsname}
\def\ifuncit@d#1{\expandafter\ifx\csname r@fnum#1\endcsname\relax}%
\def\newr@f#1{\global\advance\r@fcount by1%
    \expandafter\xdef\csname r@fnum#1\endcsname{\number\r@fcount}}

\let\r@fis=\refis			
\def\refis#1#2#3\par{\ifuncit@d{#1}
   \newr@f{#1}%
   \w@rnwrite{Reference #1=\number\r@fcount\space is not cited up to now.}\fi%
  \expandafter\gdef\csname r@ftext\csname r@fnum#1\endcsname\endcsname%
  {\writer@f#1>>#2#3\par}}

\def\ignoreuncited{
   \def\refis##1##2##3\par{\ifuncit@d{##1}%
    \else\expandafter\gdef\csname r@ftext\csname r@fnum##1\endcsname\endcsname%
     {\writer@f##1>>##2##3\par}\fi}}

\def\r@ferr{\endreferences\errmessage{I was expecting to see
\noexpand\endreferences before now;  I have inserted it here.}}
\let\r@ferences=\references
\def\references{\r@ferences\def\endmode{\r@ferr\par\endgroup}}

\let\endr@ferences=\endreferences
\def\endreferences{\r@fcurr=0
  {\loop\ifnum\r@fcurr<\r@fcount
    \advance\r@fcurr by 1\relax\expandafter\r@fis\expandafter{\number\r@fcurr}%
    \csname r@ftext\number\r@fcurr\endcsname%
  \repeat}\gdef\r@ferr{}\endr@ferences}


\let\r@fend=\endpaper\gdef\endpaper{\ifr@ffile
\immediate\write16{Cross References written on []\jobname.REF.}\fi\r@fend}

\catcode`@=12

\citeall\refto		
\citeall\ref		%
\citeall\Ref		%

\def\refto#1{$^{#1}$}         

\def\la{\langle}
\def\ra{\rangle}
\def\ria{\rightarrow}
\def\n{{\bar n}}
\def\g{{\bar g}}
\def\h{{\bar h}}

\def\x{{\bf x}}

\def\q{{\bf q}}

\def\a{\alpha}
\def\b{\beta}

\def\s{{\sigma}}
\def\e{{\epsilon}}

\def\n{{\bar n}}

\def\Tr{{\rm Tr}}
\def\ih{{i \over \hbar}}

\def\s{{\sigma}}

\def\trho{{\rho}}

\def\P{{\bar P}}
\def\p{{\partial }}

\centerline{\bf Decoherent Histories and Hydrodynamic Equations}

\author J.J.Halliwell 
\affil 
Theory Group 
Blackett Laboratory 
Imperial College 
London 
SW7 2BZ 
UK 
\vskip 0.5in 
\centerline {\rm Preprint IC/TP/97-98/50. May, 1998} 
\vskip 0.1in  
\centerline {\rm Submitted to {\sl Physical Review D}} 

\vskip 1.0in 

\abstract{ For a system consisting of a large collection of
particles, a set of variables that will generally become effectively
classical are the local densities (number, momentum, energy). That
is, in the context of the decoherent histories approach to quantum
theory, it is expected that histories of these variables will be
approximately decoherent, and that their probabilites will be
strongly peaked about hydrodynamic equations. This possibility is
explored for the case of the diffusion of the number density of a
dilute concentration of foreign particles in a fluid. This system has
the appealing feature that the microscopic dynamics of each individual
foriegn particle is readily obtained and the approach to local
equilibrium may be seen explicitly. It is shown that, for certain
physically reasonable initial states, the probabilities for
histories of number density are strongly peaked about evolution
according to the diffusion equation. Decoherence of these histories
is also shown for a class of initial states which includes
non-trivial superpositions of number density. Histories of phase
space densities are also discussed. The case of histories of number,
momentum and energy density for more general systems, such as
a dilute gas, is also discussed in outline. When the initial
state is a local equilibrium state, it is shown that the histories
are trivally decoherent, and that the probabilities for histories
are peaked about hydrodynamic equations. An argument for decoherence
of more general initial states is given. } 
\endtopmatter 
\endpage

\head{\bf 1. Introduction}

One of the aims of the decoherent histories approach to quantum
theory is to supply an account of the emergence of an approximately
classical world from an underlying quantum one
[\cite{Gri,GH1,GH2,Omn}]. To date, most of the effort in this
direction has concentrated on systems in which there is a natural
separation into distinguished system and environment. In the quantum
Brownian motion model, for example the distinguished system is a
large particle, and the environment a bath of harmonic oscillators.
Histories of imprecise positions  of the distinguished system are
then typically decoherent as a result of their interaction with the
environment, and the probabilities for histories of position are
typically peaked about classical equations of motion, with
dissipation [\cite{GH2,DoH}]. These models have the advantage of
technical simplicity -- a variety of known techniques, such as path
integral and the influence functional [\cite{FeV,CaL}],  may be
brought to bear, and emergent classicality is readily shown. These
systems have also been analysed from other perspectives,  including
the density matrix approach and quantum state diffusion approach,
with qualitatively similar results [\cite{JoZ,Zur,HaZ}].

It is however, of considerable interest to investigate models of a
more general class that might describe a gas or fluid, where there
are certainly no distinguished point particles, and possibly even no
obvious candidates for an environment to produce decoherence. In
this context it has been suggested by Gell-Mann and Hartle that for
a large and possibly complex system, the variables that will become
classical ``habitually'' are the local densities (energy, momentum
and number, for example), integrated over small volumes
[\cite{GH2}]. They are expected to be approximately decoherent
because they are approximately conserved (since exactly conserved
quantities are exactly decoherent [\cite{HLM}]). Furthermore, there
is some hope that the probabilities for such histories might be
peaked about interesting deterministic evolution equations. This is
because it is known that the local densities, averaged in a local
equilibrium state, obey hydrodynamic equations.

It is therefore of interest to study the decoherent histories
approach applied to systems consisting of large numbers of particles, with
histories consisting of projections onto local densities, with the
aim of showing, first, that these histories decohere; and second, their
probabilities are peaked about hydrodynamic equations. In this paper
we make a number of steps in this direction.

Generally, we are concerned with systems which are described at the
microscopic level by a Hamiltonian of the form
$$
H = \sum_i \left( { {\bf p}_i^2 \over 2 m_i } + \sum_{i \ne j} V_{ij}
( | \q_i - \q_j | ) \right)
\eqno(1.1)
$$
The local densities of interest are the number density
$n(\x,t)$, the momentum density ${\bf g}(\x,t)$ and the energy
density $ h (\x,t)$, defined by,
$$
\eqalignno{
n(\x,t) &= \sum_i \ \delta(\x -\q_i)
&(1.2) \cr
{\bf g}(\x,t) &= \sum_i \ {\bf p}_i \ \delta (\x- \q_i)
&(1.3) \cr
h(\x,t) &= \sum_i \ \left( { {\bf p}^2_i \over 2 m_i} + \sum_{j \ne i}
V_{ij} (| \q_i - \q_j |) \right)\ \delta(\x- \q_i)
&(1.4) \cr }
$$
When these objects are raised to the status of operators in the
quantum theory, pairs of non-commuting operators 
are symmetrized to ensure
hermiticity. (In what follows it will be assumed, where
necessary, that non-commuting operators are ordered in this way,
even  though it will not always be written explicitly). The standard
derivation of the hydrodynamic equations for the average values of
the local densities starts by considering the continuity equations 
expressing local conservation [\cite{hydro}]. These are of the form
$$
{ \partial \sigma \over \partial t} + \nabla \cdot {\bf  j } = 0
\eqno(1.5)
$$
where $\sigma $ denotes $n$, ${\bf g}$ or $h$ (and the current
${\bf j}$ is a second rank tensor in the case of ${\bf g}$).
It is assumed that, for a wide variety of initial states, 
conditions of local equilibrium  are
established after a short period of time.
This means that on scales small compared to the overall
size of the fluid, but large compared to the microscopic scale,
equilibrium conditions are reached in each local region,
characterized by a local temperature, pressure {\it etc.} which vary
slowly from region to region. This state of affairs is described by
a local equilibrium density operator, which has the form
$$
\rho = Z^{-1} \exp \left( - \int d^3x \b (\x) \left[ h(\x) - \bar \mu (\x)
n (\x) - {\bf u} (\x) \cdot {\bf g} (\x) \right] \right)
\eqno(1.6)
$$
where $ \b  $, $\bar \mu $ and $ {\bf u} $ are Lagrange multipliers and
are slowly varying functions of space and time. ($\b$ is the inverse
temperature, ${\bf u}$ is the average velocity field, and $\bar \mu$
is related to the chemical potential which in turn is related to
the average number density). The hydrodynamic
equations follow when the continuity equations are averaged in this
state. The averaging produces averages of the currents, $\la {\bf j}
\ra $, but constitutive relations then typically follow, which
relate the averages of the currents to averages of the local
densities (and their derivatives). Hence one obtains a closed set of
evolution equations for $\la n \ra $, $ \la {\bf g} \ra $ and $ \la h \ra
$ (smeared over small spatial volumes).

What is therefore known already is that the {\it average values} of
the local densities, in the local equilibrium state, obey 
hydrodynamic equations. 

Demonstrating emergent classicality, however, requires considerably
more than this. The above result is not unlike the Ehrenfest theorem
of elementary quantum mechanics. This shows that the expectation
values of position and momentum obey classical equations of motion,
but it alone does not constitute a derivation of the emergence of
classical mechanics from quantum mechanics [\cite{Har6}].

Here, we adopt the discussion of emergent classicality
used in the decoherent histories approach.
In that approach we consider the decoherence
functional,
$$
D({\underline \alpha}, {\underline \alpha}') 
= {\rm Tr} \left( P_{\a_n} (t_n) \cdots P_{\a_1} (t_1)
\rho P_{\a_1'} (t_1 ) \cdots P_{\a_n'} (t_n)  \right)
\eqno(1.7)
$$
The histories are characterized by projectors $P_{\a} (t)$ (in the
Heisenberg picture) at each moment of time, in this case onto the
local densities (the detailed construction is described below). The
decoherent histories approach is described in detail in many places,
so we will not  go into detail  here 
[\cite{Gri,GH1,GH2,Omn,DoH,Hal1}]. Briefly, the
histories are said to be decoherent when the decoherence functional
(1.7) is approximately zero in its off-diagonal terms. Probabilities
equal to the diagonal elements may then be assigned to the histories
in the set.

A decoherent histories derivation of the hydrodynamic equations 
goes beyond the standard one in two respects. First of all, it shows
that histories of local densities are decoherent, so probabilities
can be assigned to them. Secondly, rather than considering just the
evolution of average values of the local densities, we consider the
probability for histories of values, which is a far more detailed
object. We will, however, find an important simplifying feature in
the special case of a local equilibrium initial state: this is that
the probabilities for histories are strongly peaked about the
evolution of the average values. In this case, the jump from 
the evolution of average
values to probabilities for histories is then in fact not that large.

It is perhaps also of interest to note that it is difficult to
discuss the decoherence and emergent classicality  of the local
densities using other approaches to decoherence. The density matrix
approach, for example, relies quite heavily on the existence of a
reduced density matrix for the variables followed, {\it e.g.}, the
position of  Brownian particle, after the variables ignored are
traced out.  This requires that the total Hilbert space is a tensor
product of the Hilbert space of the variables followed with the
Hilbert space of the environment, which is not the case here. The
decoherent histories approach, by contrast, adapts more comfortably
to these situations. It deals with projections onto subspaces of the
total Hilbert space, and does not require a tensor product
structure.

Turning to this paper, we will address the issue of the
emergence of hydrodynamic equations for the special case of the
diffusion of local number density of foreign particles in a fluid. 
We then go on to argue that the approach and some of the
results also apply to the more general case; thus, a (partially
heuristic) derivation of the hydrodynamic equations from decoherent
histories will be given.

The system we are initially interested in is a fluid containing a
very low concentration of foreign particles. The number of foreign
particles is conserved, and it is well-known that the local number
density obeys the diffusion equation. For us, the simplifying
feature is that the microscopic dynamics of the foreign particles is
reasonably simple, and the approach to equilibrium and the
consitutive relations are easily derived. In particular, each
foreign particle may be regarded as a free particle in a bath (the
background fluid), so is essentially a particle undergoing quantum
Brownian motion, the description of which is well-known.

Hence, in this particular model there {\it is} in fact, an
environment, namely the background fluid, which assists in many
ways, in particular it makes a contribution to decoherence. What is
new, however, compared to previously considered system--environment
models, is to consider histories of local densities of many Brownian
particles, rather than histories of positions of a single point
particle. Indeed, this model is a kind of half-way house between the
familiar system--environment models and true hydrodynamic models.
Furthermore, as indicated above, the model is sufficiently similar
to the proper hydrodynamic case to make  its results of relevance.

The quantum Brownian motion model and its diffusive limit is
described in Section II. Its reinterpretation as a deterministic
equation for the number density is described in Section III.
The decoherent histories analysis is described in Section IV.
It is shown that histories are peaked about diffusive evolution,
and that they are approximately decoherent. The extent to which
these ideas go over to the more general case is discussed in Section
V. We summarize and conclude in Section VI. An alternative
proof of the key results of Sections IV and V is given in an
appendix.

A number of other authors have investigated closely related issues.
The present work is strongly rooted in the decoherent histories
approach to quantum theory of Gell-Mann and Hartle, who have
frequently stressed that the most general situations in which there
is decoherence and peaking of the probabilities about deterministic
equations are generally not of the system-environment type 
[\cite{GH1,GH2,Har6}]. An earlier paper investigated the decoherence
of histories of number density in a simple model involving spin
waves [\cite{BrH}]. Brun and Hartle have investigated coarse
grainings in an oscillator chain model which allow a wide variety of
 different divisions into system and environment, and also lead to a
wave equation as the effective classical equation of motion
[\cite{BrHa}]. Calzetta and Hu have investigated the decoherence of
histories specified by fixed values of certain correlation functions
in a field theory [\cite{CaH}]. Other coarse graining schemes not of
the system environment type (and not involving decoherent histories)
may be found in Refs.[\cite{Car,Flo}]. Some related issues
concerning the derivation of hydrodynamic equations may be found in
Ref.[\cite{Ana}]. Omn\`es [\cite{Omn2}] has observed that the work
of Feffermann [\cite{Fef}]  may contain some useful guides
concerning the selection of a preferred set of variables is a
general system.

Finally, it should be mentioned that not all authors agree that the
decoherent histories approach is capable of explaining the emergence
of classical physics from an underlying quantum theory. The various
sides of this rather subtle debate are described in
Refs.[\cite{DoK,GiR,Gri2}]. The present paper is essentially
concerned with exploring the mathematical properties of the
decoherent histories approach, as it currently exists, and has
little to add to this debate. The extent to which the decoherent
histories approach remains a valuable and useful one is perhaps best
summarized in Ref.[\cite{Har7}].

\head{\bf 2. Derivation of the Diffusion Equation}

The diffusion equation arises for a system consisting
of a collection of $N$ foreign particles in a fluid of a much larger
number of background particles. 
We assume that the density of foreign particles
is so low that their interactions with each other may be neglected.
Each particle interacts only with the fluid which we assume is in
equilibrium at temperature $T$.  Each individual foreign particle,
therefore, may be modeled as a free Brownian particle for which the
fluid plays the role of a thermal environment.  The description of
this system is well known. Classically, the phase space
density $W(p,q)$ of the particle obeys a Fokker-Planck equation of
the Kramers type, which takes the form [\cite{Ris}],
$$
{ \p W \over \p t } = - {  p \over M } { \p W \over \p q }
+ 2 \gamma { \p \over \p p } ( p W ) + 2 M \gamma k T 
{ \p^2 W \over \p p^2 }
\eqno(2.1)
$$

In the quantum theory, the Wigner function obeys the same equation
(in the high temperature limit, which we work in here),
The density operator is recovered from the Wigner function by the
transformation
$$
\rho(x,y) = \int dp \ e^{\ih p (x-y) } \ W ( p, {x+y \over 2} )
\eqno(2.2)
$$
and it obeys the master equation
$$
\eqalignno{
{ \partial  \rho \over \partial t} = 
&  { i \hbar \over 2M } \left( { \partial^2 \rho \over \partial x^2 }
- { \partial^2 \rho \over \partial y^2 } \right)
\cr & 
-  \gamma  (x-y) \left( { \partial \rho \over \partial x}
- { \partial  \rho \over \partial y} \right)
-  {2 M \gamma k T \over \hbar^2 } (x-y)^2 \trho 
&(2.3) \cr }
$$
This equation arises directly in the Caldeira-Leggett model of
quantum Brownian motion, in which a particle is linearly coupled to
a thermal bath of harmonic oscillators [\cite{CaL}]. However, there
are many indications [\cite{Omn3,Dio}] that the
form of this equation is in fact a lot more general, and
will still hold, in some approximation, for more complicated and
realistic couplings.

Eq.(2.1) may be solved by introducing the
propagator $K (p,q,t|p_0,q_0,0) $, defined by
$$
W(p,q,t) = \int dp_0 dq_0 \ K(p,q,t | p_0, q_0, 0) 
\ W_0 (p,q)
\eqno(2.4)
$$
The propagator also obeys Eq.(2.1)
with the initial conditions,
$$
K(p,q,0|p_0,q_0,0) = \delta (p-p_0) \ \delta (q-q_0)
\eqno(2.5)
$$
The explicit form is,
$$
K(p,q,t  |p_0,q_0,0)  = 
\ \exp \left( - \a (p - p^{cl} )^2 -\b (q- q^{cl})^2 -
\epsilon (p-p^{cl})(q-q^{cl}) \right)
\eqno(2.6) 
$$
(ignoring prefactors, which may be recovered by normalization where necessary).
Here,
$$
\eqalignno{
q^{cl} &= q_0 + { p_0 \over 2 M \gamma} ( 1 - e^{ -2 \gamma t } )
\cr
p^{cl} &=p_0 e^{ - 2 \gamma t }
&(2.7) \cr }
$$
Explicit expressions for the coefficients $\a$, $\b$, $\e$ may
be found in Ref.[\cite{AnH}].

We are interested in the behaviour of this system for times
$ t >> \gamma^{-1} $. Unlike the harmonic oscillator, this does not
settle down to equilibrium, but certain simplifications do occur.
First of all, note that
$$
\eqalignno{
q^{cl} \ria & \ q_0 + { p_0 \over 2 M \gamma}
\cr
p^{cl} \ria & \ 0
&(2.8) \cr }
$$
Second, from the explicit expressions in Ref.[\cite{AnH}], it can
be shown that
$$
\a \ria {1 \over 2 M k t }, \quad
\b \ria { M \gamma \over 2 k T t}, \quad 
\e \ria - { 1 \over 2 k T t }
\eqno(2.9)
$$
The solution Eq.(2.1) therefore approaches the form,
$$
W(p,q,t) = \int dp_0 dq_0 
\ \exp \left( - \a p^2 -\b (q- q^{cl})^2 -
\e p (q-q^{cl}) \right)
\ W_0(p,q)
\eqno(2.10)
$$
The distribution of momenta is
$$
g(p) = \int dq \ W (p,q,t) 
 \approx \exp \left(- { p^2 \over 2 M k T } \right)
\eqno(2.11) 
$$
(where we have used $ \gamma t >> 1 $). Hence the momentum
distribution approaches the Maxwell-Boltzmann form. The
distribution of positions, on the other hand, approaches
the form
$$
\eqalignno{
f (q,t) &= \int dp \ W (p,q,t)
\cr
& \approx \int dp_0 dq_0 \exp \left(
- { M \gamma \over 2 K T t } \left( q - q_0 - { p_0 \over 2 M
\gamma} \right)^2 \right) \ W_0 (p_0, q_0 )
&(2.12) \cr }
$$
It follows that $ f(q,t)$ obeys the diffusion equation,
$$
{ \p f \over \p t } = 
D { \p^2 f \over \p q^2 }
\eqno(2.13)
$$
where $ D = { k T / 2 M \gamma} $.

This equation can also be derived by integrating Eq.(2.1) over $p$,
with the result,
$$
{ \p f \over \p t } = - { 1 \over M } { \p \over \p q }
\int dp \ p \ W (p,q,t)
\eqno(2.14)
$$
and Eq.(2.10) may be then be used to show that,
$$
\int dp \ p \ W (p,q,t) =
- { k T \over 2 \gamma} { \p \over \p q } f(q)
\eqno(2.15)
$$
hence the diffusion equation follows.



\head{\bf 3. Diffusion of Number Density}

Turn now from the case of a single particle to the system of
interest --
a collection of $N$
particles, each one of which is described by the one-particle
dynamics outlined above. The number
density at the point $x$ is
$$
n(x) = \sum_{k=1}^N \ \delta (q_k - x)
\eqno(3.1)
$$
Usually, we are interested in the number density
smeared over a small volume $V$, which we write,
$$
n_V (x) = \sum_{k=1}^N \ \delta_V (q_k - x)
\eqno(3.2)
$$
where $\delta_V $ denotes a delta function smeared over $V$, 
{\it i.e.},
a window function equal to $1$ in an interval of size $V$ 
and $0$ outside, and $x$ will then be a discrete label.
(We will not explicitly use the smeared form unless necessary). 
Microscopically, we assume that the $N$ particle system is described
by the phase space distribution function
$$
\mu ( p_1, p_2, \cdots p_N , q_1, q_2, \cdots q_N) =  
\prod_{k=1}^N \ W (p_k, q_k, t )
\eqno(3.3)
$$
where $W(p,q,t)$ are the one-particle Wigner functions of
the previous section. The product form is reasonable since
the particles do not interact, and even if they are initially
correlated, these correlation will eventually be lost
to the environment of each particle.

The number density $n(x)$ is a function of the stochastic variables
$q_k$. However, we will show that the evolution of $n(x)$ is in fact
strongly peaked about deterministic evolution according to the
diffusion equation. This is, of course, almost obvious, since all we
are doing is in essence reinterpreting the one-particle probability
as a distribution function on a real ensemble of $N$ particles. But
for what follows, it is worth going through the steps. First we show
that the average value obeys the diffusion equation, then we show
that the distribution is strongly peaked about the average.

Consider first the average of $n(x)$ in the $N$ particle
distribution (3.3). It is,
$$
\eqalignno{
\la n (x) \ra & = \sum_k \ \la \delta (q_k - x ) \ra_N 
\cr
&= N \la \delta (q -x ) \ra_1 
\cr
&= N \int dp W (p,x,t ) = N f(x)
&(3.4) \cr }
$$
where $ \la \cdots \ra_N $ and $ \la \cdots \ra_1 $ denote
averages in the $N$-particle and one-particle distributions,
respectively.
Hence $ \la n (x) \ra $ obeys the diffusion equation, since $ f(x)$
does, as shown in the previous section.

It is, however, enlightening to briefly demonstrate this a different
way, more closely related to standard derivations of hydrodynamics
equations. The number density obeys the local conservation law,
$$
{ \p n \over \p t } + {1 \over M} { \p g \over \p x } = 0
\eqno(3.5)
$$
where
$$
g (x) = \sum_k \ p_k  \ \delta ( q_k - x )
\eqno(3.6)
$$
Averaging through (3.5), we find we need to compute 
$$
\eqalignno{
\la g (x) \ra  &= \sum_k 
\ \la  p_k  \ \delta ( q_k - x ) \ra_N
\cr
& = N \int dp \ p  
\ W (p,q,t )
&(3.7) \cr }
$$
We have already computed this in Eq.(2.15). Hence we derive the
constitutive relation,
$$
\la g (x) \ra = - { k T \over 2  \gamma } 
\ {\p \over \p x} \la n ( x) \ra  
\eqno(3.8)
$$
and the diffusion equation for $ \la n (x) \ra $ follows. Note also
that the state averaged in here is close to a local equilibrium
state -- the momenta are in a Maxwell-Boltzmann distribution
(2.11), the temperature $T$ is constant throughout the system,
the average velocity flow is zero, and only the number density
varies throughout the system (although the correlation between
$p$ and $q$ in (2.10) is important for (3.8) to be obtained).

Now consider the variance of $n(x)$ in the distribution (3.3).
Here we have to consider the smeared from (3.2). We have
$$
\la n_V  \ra = \sum_k \ \la \delta_V (q_k - x) \ra_N 
= N \la \delta_V (q - x ) \ra_1
\eqno(3.9)
$$
and
$$
\eqalignno{
\la n^2_V \ra &= \sum_{jk} \la \delta_V (q_j -x ) \delta_V (q_k - x )
\ra_N 
\cr
&= \sum_j \la \delta_V^2 (q_j - x ) \ra_N + \sum_{j \ne k}
\la \delta_V (q_j -x ) \ra_N \ \la  \delta_V (q_k - x ) \ra_N
\cr
&= N \la \delta_V^2 (q-x) \ra_1 + (N^2 - N) \la \delta_V (q-x)
\ra_1^2
&(3.10) \cr }
$$
Therefore,
$$
{ \la n^2_V \ra - \la n_V \ra^2 \over \la n_V \ra^2 } = {1 \over N }
{ \la \delta_V^2 \ra - \la \delta_V \ra^2 \over \la \delta_V \ra^2 }
\eqno(3.11)
$$
For large $N$, which we assume, the distribution of $n_V (x)$ is very
strongly peaked about its mean value.

The above result, which is a form of the central limit theorem, is
important in the derivation of more general hydrodynamic equations
(and indeed is used again in the next section in a slightly
different form). It essentially relies on two  features of the
macroscopic system. Firstly, that the microscopic distrbution is of
the form (3.3), {\it i.e.}, a product of identical one-particle
distribution functions. Secondly, that the macroscopic variable in
question, such as (3.1), is a sum of identical terms each of which
involves only the one-particle variables. It is, however, very
likely (and probably provable) that the result will hold more
generally. In particular, we expect it to continue to hold in the
more realistic situation where the microscopic distribution function
involves a certain amount of correlation between particles (so is
not exactly of the product form) and when the macroscopic variables
involve interaction terms between particles. Generally, we expect
strong peaking about the mean values of a macroscopic variable when
fixed values of that variable correspond to a very large number of
similar microstates.



\head{\bf 4. Histories of Number Densities}

\subhead{\bf 4(A). Probabilities and Decoherence}

Consider now the question of histories of number densities.
After tracing out the environment for each of the $N$ particles,
the decoherence functional for histories characterized by projections
onto number density at two moments of time is
$$
D (\n_1 , \n_2 | \n_1', \n_2 ) = {\rm Tr} \left( P_{\n_2} K_0^t \left[
P_{\n_1} \rho P_{\n_1'} \right] \right)
\eqno(4.1)
$$
The trace is over the Hilbert space of the $N$ particles.  
$K_0^t$ is a product of $N$ reduced density
matrix propagators (corresponding to the master equation (2.3)), 
one for each Brownian particle. $P_{\n}$ denotes a
projector onto number density, and may be represented as
$$
P_{\n (x) } = \delta_{\s} \left[ \hat n (x) - \bar n (x) \right]
\eqno(4.2)
$$
Here, $ \delta_{\s} $ is a delta function smeared over a range $\s$.
The number density operator $\hat n(x) $ may be smeared over a range
$V$ when necessary. (More detailed representations of projections
onto number density are possible, in terms of projections onto
position of the individual particls [\cite{BrH}], but this is not
necessary here). Eq.(4.1) is the decoherence functional for
the simplest non-trivial case of
two-times histories, but the arguments present below will also go
through for histories of projectors at many moments of time.

For the initial state $\rho$, let us again suppose that it is a
product of $N$ identical terms, as in the previous section. We will
also assume that it has  already evolved for a time $t >>
\gamma^{-1}$, so we are in the diffusive regime. We will consider
later the consequences of other types of initial state.

If the initial density
matrix is a product of $N$ identical terms, we expect a central
limit theorem result to hold, and the distribution of $n(x)$ in the
initial  state will be strongly peaked about its mean. 
In more detail, suppose we have a pure initial state on the $N$ particle
Hilbert space which is a product of $N$ identical terms,
$$
| \Psi \ra = | \psi \ra \otimes | \psi \ra \otimes \cdots \otimes |
\psi \ra
\eqno(4.3)
$$
Then for large $N$, $ | \Psi \ra $ is approximately 
an eigenstate of the number density operator (2.1),
$$
\hat n(x) | \Psi \ra \ \approx \  \la n(x) \ra | \Psi \ra
\eqno(4.4)
$$
where $ \la n(x) \ra = N | \psi (x) |^2 $. This is essentially 
the Finkelstein--Graham--Hartle theorem [\cite{FGH}], since the
number density is the same as the relative frequency
operator (up to an overall factor of $N$). 
For projections onto number density a similar result
follows, that is,
$$
\delta_{\s} \left[ \hat n(x) - \n (x) \right] | \Psi \ra
\ \approx \ \delta_{\s} \left[ \la n(x) \ra - \n (x) \right] | \Psi \ra
\eqno(4.5)
$$
It is straightforward to
generalise to the
case of mixed states. That is, for a density matrix
$$
\rho = \rho_1 \otimes \rho_1 \otimes \cdots \otimes \rho_1
\eqno(4.6)
$$
it may be shown that
$$
\hat n(x) \rho \approx \la n(x) \ra \rho
\eqno(4.7)
$$
and thus, the analogue of (4.5) holds.

These results mean two things.
First, that 
$$
P_{\n_1} \rho P_{\n_1'} \approx 0, \quad {\rm for} \quad \n_1 \ne
\n_1 '
\eqno(4.8)
$$ 
Hence, there is approximate decoherence for initial states of this type.
Secondly, they allow us to compute the probabilities. We have
$$
P_{\n_1} \rho P_{\n_1} \approx \ \delta_{\s} \left[ \la \hat n (x) \ra
- \n_1 (x) \right] \rho 
\eqno(4.9)
$$
Inserting into the diagonal elements of the decoherence functional
(4.1), we find
$$
p( \n_2, \n_1 ) = {\rm Tr} \left( P_{\n_2} \rho_t \right)
\ \delta_{\s} \left[ \la \hat n (x) \ra - \n_1 (x) \right]  
\eqno(4.10)
$$
Now $\rho_t $ is the evolution of the initial state under the
non-unitary propagator $K_0^t $.
The key point is that it remains in the product
form, so,
$$
{\rm Tr} \left( P_{\n_2} \rho_t \right) \approx
\ \delta_{\s} \left[ \la \hat n (x) \ra_t - \n_2 (x) \right]  
\eqno(4.11)
$$
Hence, the probability (4.10) is strongly peaked about
$ \n_1 (x) = \la \hat n (x) \ra $ and $\n_2 (x) = \la \hat n(x)
\ra_t $, that is, about evolution according to the diffusion
equation.

We have therefore proved the desired result: for initial states of
the form (4.6), histories of number density are decoherent, and
their probabilities are peaked about diffusive evolution.

\subhead{\bf 4(B). More about Decoherence}

Let us consider now in more detail the question of decoherence.
Since each individual foreign particle is coupled to its surrounding
fluid environment, the possibility exists that this environment 
causes decoherence of each particle's position. Decoherence of
number density then follows trivially, since projections onto number
density may be constructed from coarse-grainings of projections onto
position [\cite{BrH}]. However, decoherence of a particle coupled to
an environment requires certain choices of parameters for the model,
such as large mass and large temperature. Hence, there will be
certain parameter ranges (in particular, small masses for the
foreign particles) for which the decoherence at the one particle
level may not be very good.  In this case, it is the large $N$
effect which produces the dominant contribution to decoherence of
number density. (Generally, replicating the system $N$ times
enhances the degree of decoherence exponentially in $N$
[\cite{BrH}]).

Because decoherence is not necessarily produced by the environment
of each particle, it is of interest to see how decoherence comes
about for initial states consisting of superpositions of number
density. So let,
$$
\eqalignno{
| \Psi \ra = { 1 \over \sqrt{2} } & \left(
| \psi \ra \otimes | \psi \ra \otimes \cdots \otimes | \psi \ra \right.
\cr
& \quad \quad \left. + | \chi \ra 
\otimes | \chi \ra \otimes \cdots \otimes | \chi \ra \right)
&(4.12) \cr}
$$
It is now the case that $P_{\n_1} \rho P_{\n_1'} \ne 0 $
if we take $\n_1 (x) = N | \psi (x) |^2 $ and 
$\n_1' (x) = N | \chi (x) |^2$. 

In this case, decoherence may be demonstrated by appealing to
approximate determinism: decoherence follows if the probabilities
are strongly peaked about a unique relationship between alternatives
at different times.  To see this, note that for histories consisting
of alternatives at two moments of time, the decoherence functional
satisfies [\cite{DoH}]
$$
| D( \n_1, \n_2 | \n_1', \n_2 ) |^2 \le
p (\n_1, \n_2 ) p (\n_1', \n_2 )
\eqno(4.13)
$$
So if $ p (\n_1, \n_2 ) $ is strongly peaked about an equation
linking $ \n_1 $ and $\n_2 $, only one of $ p(\n_1, \n_2) $
and $ p (\n_1', \n_2) $ can be non-zero for $\n_1 \ne \n_1' $,
hence the decoherence functional is approximately diagonal.

Before looking in more detail at how this argument applies to the
specific case of diffusive evolution, some qualifications must be
added. For diffusive evolution there is not, in fact, a unique
initial density corresponding to a fixed final one.  There will
generally be a spread of initial densities that evolve into a given
final density, and the non-uniqueness will get worse with increasing
time intervals. A fixed final density will fix the initial density
precisely only for reasonably short times, and for sufficiently
large choices of the width $\s$ of the projections onto number
density (which ensures that different values of $n(x)$ are
sufficiently ``far apart"). Fortunately, these restrictions do not
adversely affect the main objective here -- to demonstrate the
emergence of the diffusion equation -- since it is clearly
sufficient to consider projectors onto number density at two closely
separated times.

Returning to the demonstration of decoherence, the probabilities
$p(\n_1, \n_2 )$ are non-zero only if $ \n_1 $ is equal to either $
N | \psi (x) |^2 $ or $N | \chi (x) |^2 $. The first choice projects
the initial state onto $ |\psi \ra \otimes | \psi \ra \otimes \cdots
$ and the second onto $ |\chi \ra \otimes | \chi \ra \otimes \cdots
$. The point is now that each of these two states, on evolution for
a short time $t$, are strongly peaked about two distinct values of
number density, as determined via Eq.(2.12). (It is not necessarily
diffusive evolution, except for times $t >> \gamma^{-1} $. It is
only necessary that the final value of number density be uniquely
determined by the initial state).  It follows that, modulo the above
remarks, $p(\n_1, \n_2)$ is strongly peaked about a unique relation
between $\n_1 $ and $\n_2$, hence there is approximate  decoherence.

\subhead{\bf 4(C). Other Variables}

It is possible to consider histories of other types of variables in
this diffusive model. Momentum and energy are not so interesting
because they are not locally conserved -- the momentum and energy of
the foreign particles is exchanged with the environment. But another
finer-grained variable that is likely to be of more general use is
the phase space density. Classically, this is given by
$$
f(x,k) = \sum_j \delta (p_j - k ) \delta (q_j - x )
\eqno(4.14)
$$
This quantity may be smeared over ranges of $p$ and $q$, so that
it becomes a function $ f_{\Gamma} (x,k)$ which counts the number of
particles in the phase space cell of size $\Gamma$ labeled by
discrete lables $x,k$. Classically,
$$
\la f(x,k) \ra_N = N W (k,x) 
\eqno(4.15)
$$
where $W(k,x)$ is the one-particle phase space distribution
function. Like the number density, the distribution of
$ f(x,k)$ is strongly peaked about its mean, (4.15), since (4.14)
is a sum of identical terms. Hence, we can regard the phase space
density (4.14) as a hydrodynamic variable whose evolution equation
is the Fokker-Planck equation (2.1).

Similar results will hold in the quantum theory.
Operator ordering is necessary to make (4.14) hermitian.
We can the consider projections onto ranges of its values,
$$
\delta_{\s} \left[ \hat f_{\Gamma} (x,k) - \bar f (x,k) \right]
\eqno(4.16)
$$
The subsequent treatment will be similar to the number density case.
One would expect the quantum theory to exhibit an approximate
determinism in the evolution of (4.14) (similar to Omn\`es theorems
on the evolution of phase space projectors [\cite{Omn}])  and this
will guarantee approximate decoherence of histories of phase space
density. The probabilities for these histories will then be peaked
about evolution according to the Fokker-Planck equation.
(This will be investigated in more detail elsewhere).

\subhead{\bf 5. General Hydrodynamic Histories}

Interestingly, many of the features of the diffusive model
go over to the general case.
Consider
a closed system consisting of $N$ weakly interacting particles,
now {\it without} a background fluid.
The number, momentum, and energy
density are given by 
$$
\eqalignno{
n(x) &= \sum_j \delta (q_j -x )
&(5.1) \cr
g(x) &= \sum_j \delta (q_j - x ) p_j
&(5.2) \cr
h(x) & = \sum_j \delta (q_j - x) { p_j^2 \over 2m }
&(5.3) \cr }
$$ 
They are written for the case of a one-dimensional model for
notational simplicity, although the three-dimensional case is
readily obtained. The Hamiltonian is for a system of approximately
non-interacting particles. This is sufficient to cover, for example,
the case of a dilute gas. The neglect of interactions might seem
restrictive, but, as in the diffusive model, we find that  the
interesting things we are able to say about the probabilities
histories arise from large $N$ statistics, rather than from detailed
knowledge of the microscopic interactions.

As in the previous case, the densities need to be smeared over a
small volume. Also, in the quantum case, non-commuting operators
should be ordered in such a way that they are hermitian.

\subhead{\bf 5(A). Decoherence and Probabilities for a Local
Equilibrium Initial State}

The usual derivation of the hydrodynamic equations [\cite{hydro}]
involves assuming a local equilibrium state, which for this
one-dimensional case has the form
$$
\rho = Z^{-1} \exp \left( - \int dx \b (x) \left[ \hat h(x) - \bar \mu (x)
\hat n (x) - u (x) \hat g (x) \right] \right)
\eqno(5.4)
$$
Because of the delta-functions in (5.1)--(5.3) this becomes,
$$
\rho = Z^{-1} \exp \left( - \sum_j  \b ( q_j) \left[ {p_j^2 \over 2
m }  - \bar \mu ( q_j ) - u ( q_j ) p_j \right] \right)
\eqno(5.5)
$$
The usual manipulations on this object may be performed by assuming
that the Lagrange multipliers $\b$, $\bar \mu$, $u$ are slowly
varying functions. In particular, one may show that the 
corresponding one-particle
Wigner function is 
$$
W_1 (p,q) = f(q) \exp \left( - { (p - m u )^2 \over 2 m k T (q) }
\right)
\eqno(5.6)
$$
where $ k T = 1 / \b $.
The quantities $f(q)$, $u$ and
$T$ are assumed to be slowly varying functions of $q$ and $t$.
(The quantity $f(q)$ is proportional to $ e^{\b  \mu} $,
where $\mu = \bar \mu - \half m u^2 $).
It is straightforwardly shown 
that the averages in this state,
$ \la n(x) \ra $, $ \la g(x) \ra $, $ \la h(x) \ra $ obey 
a simple set of hydrodynamic equations.

In the quantum theory, projectors onto local densities are again
constructed using delta-functions. The projection onto number
density is as before Eq.(4.2) and the projectors onto momentum
and energy density are
$$
\eqalignno{
P_{\g} &= \delta_{\s_g} \left[ \hat g(x) - \g (x) \right]
&(5.7) \cr
P_{\h} &= \delta_{\s_h} \left[ \hat h(x) - \h (x) \right]
&(5.8) \cr }
$$
Again $\delta_{\s}$ denotes a delta function smeared over a range
$\s$. Strictly, there is one such delta-function for each value of
$x$ (which is discrete if $\delta ( q_j - x) $ is smeared over a
small range).  Although the operators $ \hat n(x) $, $ \hat g(x) $,
$ \hat h(x) $ do not commute, the projections onto ranges of their
values $ P_{\n} $, $ P_{\bar g} $, $ P_{\bar h} $ will approximately
commute with each other if both the width $\s$ of the projections
and the smearing volume of $\delta (q_j -x ) $ is sufficiently
large. (The extent to which this approximation is good will
also depend to some extent on the initial state they are traced with
in the decoherence functional).

In the approximation that these projectors commute,
the decoherence functional for a two time hydrodynamic history are
$$
D(\a_1, \a_2 | \a', \a_2 ) = {\rm Tr} \left( P_{\n_2} P_{\g_2} P_{\h_2}
e^{- i H t } P_{\n_1} P_{\g_1} P_{\h_1}
\ \rho \ P_{\n_1} P_{\g_1} P_{\h_1} e^{ i H t } \right)
\eqno(5.9)
$$
We take the initial state $\rho$ to be
the local equilibrium state (5.4). 
From what has gone before, it should be clear that
results similar to (4.8), (4.9) in the diffusive model hold.
That is, if
the widths of the projections $\s_n $, $\s_g $, $\s_h $ are much
greater than the variances of the corresponding operators in the
local equilibrium state, we have
$$
P_{\n_1} P_{\g_1} P_{\h_1}
\ \rho P_{\n_1'} P_{\g_1'} P_{\h_1'} \approx 0
\eqno(5.10)
$$
for $\n_1 \ne \n_1'$, $\g_1 \ne \g_1'$, $\h_1 \ne \h_1'$. Hence
there is decoherence for this initial state.
Similarly,
$$
\eqalignno{
\P_{\n_1} P_{\g_1} P_{\h_1}
\ \rho P_{\n_1} P_{\g_1} P_{\h_1} \approx
& \ \delta_{\s_n} \left[ \la \hat n(x) \ra - \n_1 (x) \right]
\cr
& \times
\ \delta_{\s_g} \left[ \la \hat g(x) \ra - \g_1 (x) \right]
\ \delta_{\s_h} \left[ \la \hat h(x) \ra - \h_1 (x) \right]
\rho
&(5.10) \cr}
$$

These two results follow from the fact that the local equilibrium
state is very strongly peaked in the locally conserved quantities
about their average values for large $N$. The proof of this property
of the equilibrium state is similar to the proof, in Section 3, that
the distribution of number density in the diffusive model is
strongly peaked about its mean. That proof relied on the property
that the averages of $n_V$ and $n_V^2$ are sums of {\it identical}
terms involving averages in the one-particle distribution functions
(see Eqs.(3.9)--(3.11)). In the local equilibrium state,  the
averages of functions of the local densities are not exactly sums of
identical terms, since the parameters of the distribution, such as
$\b$, are not constant. That is, the one-particle distribution
functions such as (5.6) are not quite the same from one particle to
another. The required peaking about the mean will still hold,
however, because the parameters of the local equilibrium state are
slowly varying functions.

After being hit by the initial projection, the density matrix is
evolved to time $t$, where it is still a local equilibrium state,
and hitting it again with a projection yields the average values at
time $t$. Hence,
$$
\eqalignno{
p(\a_1, \a_2 )  \approx 
& \ \delta_{\s_n} \left[ \la \hat n(x) \ra_t - \n_2 (x) \right]
\ \delta_{\s_g} \left[ \la \hat g(x) \ra_t - \g_2 (x) \right]
\ \delta_{\s_h} \left[ \la \hat h(x) \ra_t - \h_2 (x) \right]
\cr
\times & \delta_{\s_n} \left[ \la \hat n(x) \ra - \n_1 (x) \right]
\ \delta_{\s_g} \left[ \la \hat g(x) \ra - \g_1 (x) \right]
\ \delta_{\s_h} \left[ \la \hat h(x) \ra - \h_1 (x) \right]
&(5.12)\cr }
$$
The probability distribution is therefore peaked about the
hydrodynamic equations. (An alternative derivation of this result,
using Gaussian projections, is given in Appendix A).

This is the main result of this section: for a local equilibrium
initial condition, histories of local densities are approximately
decoherent and their probabilities are peaked about evolution
according to the hydrodynamic equations.

The discussion of finer-grained histories involving projections onto
phase space density, Eqs.(4.14)--(4.16) also applies to the more
general case. Here, the phase space density is not unlike a locally
conserved quantity, since it can change only as a result of a net
flow in phase space over the boundary of the local phase space cell.
(This is not the case in the diffusive model). The main difference
is that in the general case we would expect the one-particle phase
space density to obey a Boltzmann equation, rather than a
Fokker-Planck equation (2.1).

\subhead{\bf 5(B). Decoherence More Generally}

Turn now to the question of decoherence for more general initial
states. As stated in the Introduction, we expect the approximate
conservation of the local densities to ensure their approximate
decoherence. This is because exactly conserved quantities are
exactly decoherent [\cite{HLM}], which may be seen as follows.
Consider the two-time decoherence functional:
$$
D(\a_1, \a_2, | \a_1', \a_2 ) = {\rm Tr} \left( P_{\a_2} e^{- i H t}
P_{\a_1} \rho P_{\a_1'} e^{ i H t } \right)
\eqno(5.13)
$$
Suppose the projections are onto some exactly conserved quantity.
Then $P_{\a_2}$ commutes with $H$, so it may be moved through the
unitary evolution operator to act directly on $P_{\a_1}$ or
$P_{\a_1'}$. Since $P_{\a_2} P_{\a_1} = 0 $ unless $ \a_2 = \a_1 $,
we see that the decoherence functional is zero unless $ \a_1 = \a_2 =
\a_1'$, hence there is decoherence, for every initial state.

More generally, there will also be (approximate) decoherence if the
quantum theory exhibits some kind of approximate determinism. 
Suppose that evolution of the projectors is such that $ e^{ i H
t } P_{\a_2} e^{ - i H t } \approx P_{\a_2 (t)} $. That is, it is
approximately equal to another projector of the same operator (but onto
a different range, labeled by $\a_2 (t) $). Then again the
decoherence functional is approximately zero unless $\a_1 = \a_2 (t)
= \a_1' $. An example of this type of decoherence may be found in
the histories of phase space projectors, considered by Omn\`es. He
showed that a projector onto a sufficiently large cell $\Gamma_0 $
of phase space evolves under unitary evolution into another phase
space projector onto the phase space cell $\Gamma_t $, the classical
evolution of $\Gamma_0$.

In a similar way, we can argue for the decoherence of local
densities for a wide class of initial states by considering the time
evolution of the final projector.
Take it to be of the form
$$
P_{\a_2} =
\delta_{\s_n} \left[  \hat n(x)   - \n_2 (x) \right]
\ \delta_{\s_g} \left[  \hat g(x)  - \g_2 (x) \right]
\ \delta_{\s_h} \left[ \hat h(x)  - \h_2 (x) \right]
\eqno(5.14)
$$
and similarly for the initial projectors, $P_{\a_1}$. 
The question of decoherence
is the question of the general form of the object,
$$
P_{\a_2} (t) = e^{ i H t} P_{\a_2} e^{- i H t}
\eqno(5.15)
$$
In particular, if this object is strongly localized in
$ n $, $g$ and $ h$, then there will be approximate decoherence,
since
$$
P_{\a_1'} P_{\a_2} (t) P_{\a_1} \approx 0, \quad {\rm for } \quad
\a_1 \ne \a_1' 
\eqno(5.16)
$$
This is of course an operator statement, so involves some kind of
restriction on the class of initial states $\rho$ it is traced with
in the decoherence functional. But we would expect it to be true for
a wide class of states, although clearly not all.

To see why (5.16) should hold, regard the final projector as
representing a final state, 
$\rho_f = P_{\a_2} / {\rm Tr} (P_{\a_2} ) $ 
(the trace is finite). So Eq.(5.15) is the backwards evolution
of this state,
$$
\rho_f (-t ) = e^{ i H t} \rho_f e^{- i H t}
\eqno(5.17)
$$
Now, for unitary evolution, evolution backwards in time has the same
generic physical features as evolution forwards in time. In
particular, it is strongly believed that a wide class of initial
conditions for a large many particle system will quickly evolve to a
local equilibrium state [\cite{eqm}]. Therefore, we expect that for
sufficiently large $t$ (5.17) will be close to a local equilibrium
state, hence will be localized in $n$, $g$ and $h$, and so (5.16)
will hold. 

The issue is further helped along by the observation that $P_{\a_2}$
and hence $\rho_f $ is initially localized in $n$, $g$ and $h$, and,
since these are locally conserved quantities, both they and their
fluctuations will vary slowly.
In fact, $\rho_f$ is closely related to a local equilibrium state.
It is in a sense the microcanonical version of it. In $\rho_f$,  the
variables $n$, $g$ and $h$ are constrained to lie in a small range
centred about $\n$, $\g$ and $\h$. In the local equilibrium state,
on the other hand, only the averages of $n$, $g$ and $h$ are
constrained to take set values, but large $N$ statistics guarantees
that the fluctuations about the mean are extremely small, hence the
distributions of $n$, $g$ and $h$ are here also concentrated in a
small range centred about the mean.

Since $\rho_f$ is either essentially equivalent to a local
equilibrium state, or very close to one, and since a local
equilibrium state remains so under unitary evolution, it is very
plausible that (5.17) is localized in the hydrodynamic variables
(and in particular it is probably a local equilibrium state), so
(5.16) will hold. We therefore expect decoherence for a wide variety
of initial states.

It is perhaps pertinent to elaborate a little on what it means for a
state to approach the local equilibrium form (5.4). We expect, for
instance, a physical system starting out in a pure initial state to
approach local equilibrium, in some sense.
Yet (5.4) is a very mixed state so
cannot be reached from a pure state by unitary evolution. What is
going on here is that one is typically interested only in the
averages of the local densities, and for these purposes, the quantum
state of the system (which may be pure) can be replaced with another
state possessing the same average values. Usually, the replacement
state is chosen by the maximum entropy method (subject to fixed
average values), and it is this which leads to the local equilibrium
state (5.4) [\cite{Zub}]. 

Differently viewed, since the local densities are sums of one-particle
operators, their averages involve only the one-particle
distributions functions, such as (5.6). (If interactions are
significant two-particle operators and distribution functions could
be involved). These distribution functions may exhibit the local
equilibrium form (5.6), even though the actual state of the whole
system is not of this type. The local equilibrium density
operator of the whole system, (5.4), is the effective state of the
whole system given that the one-particle distribution functions are
of the form (5.6).

In this connection, it is also worth mentioning a different argument
for decoherence. We are interested in histories of projections onto
number, momentum and energy density on the system of $N$ particles.
However, as indicated in Ref.[\cite{BrH}], these projectors may be
expressed in terms of a (generally rather complicated) sum of
projections involving each individual particle. Hence, we could
compute the decoherence functional for the local densities by first
considering the decoherence functional for the much finer-grained
histories involving the position, momentum density and energy
density of each particle. Now the point is that the decoherence
functional for each individual particle would then have the familiar
system--environment form, in which the properties of one or two
particles are followed and all the rest are traced out. On general
grounds, we would therefore expect a degree of decoherence of the
properties of each individual particle. Since the local densities of
the $N$ particle system are constructed from coarse grainings of the
individual particle histories, the degree of decoherence can only
improve (as seen explicitly in Ref.[\cite{BrH}]).
From this point of view, the hydrodynamic models are in
fact very similar to the diffusive model of this paper, except that
here each particle doubles up as both system and environment.

The main difference, however, between the hydrodynamic models and
the usual system--environment models, is that there are no
distinguished particles, since they will typically all have the same
mass. So although there may be a degree of decoherence  in the
one-particle reduced density matrices, unlike the quantum Brownian
motion models, there will be little determinism in the evolution of
the single particle positions and momenta, because the fluctuations
will be very large as a result of interacting with an "environment"
consisting of large numbers of particle of the same mass as the
``system''. This, in fact, is one of the reasons it is interesting
of consider instead histories of local densities of the $N$ particle
system -- averaging over very large numbers of particles smooths out
the fluctuations, so that the local densities evolve almost
deterministically.

\head{\bf 6. Summary and Conclusions}

For a local equilibrium initial state, the derivation of
hydrodynamic equations from decoherent histories is reasonably
simple. Large $N$ statistics guarantees that the probabilities for
histories of local densities are strongly peaked about a single
history -- the evolution of the mean in the local equilibrium
initial state. We therefore readily connect with  previous work,
which shows that the mean values of local densities evolve
according to the hydrodynamic equations. Furthermore, the fact that
there is essentially only one history with non-zero probabilty means
that there is approximate decoherence.

Going beyond the assumption of a local equilibrium state, is
however, much more difficult. We were able to see how this sort of
state arises asymptotically in the simple diffusive model, but it is
difficult to demonstrate in general. We gave some arguments for
decoherence for more general initial states, but these still remain
rather heuristic, and it would be useful to develop some explicit
models in which decoherence via approximate conservation could be
seen more clearly. This will be addressed in future publications.

\head{\bf Acknowledgements}

I am deeply grateful to Todd Brun and Jim Hartle. Much of this work
arose from many conversations with them over a long period of time.

\def\refto#1{$^{#1}$}         

\def\la{\langle}
\def\ra{\rangle}
\def\ria{\rightarrow}
\def\n{{\bar n}}

\def\x{{\bf x}}

\def\a{\alpha}
\def\b{\beta}

\def\s{{\sigma}}
\def\e{{\epsilon}}

\def\Tr{{\rm Tr}}
\def\ih{i}

\def\s{{\sigma}}

\def\trho{{\rho}}

\def\P{{\bar P}}
\def\p{{\partial }}

\head{\bf Appendix A: An Ehrenfest Theorem for Histories}

In this appendix, we give an alternative derivation of the result of
Section V, that the probabilities for histories of local densities
are peaked about the evolution of the mean values (and hence about
the hydrodynamic equations). There, the result relied on large $N$
statistics. Here, we show that it may be regarded as an example of a
simple theorem for histories similar to the Ehrenfest theorem of
elementary quantum mechanics. The result is as follows: for any
Hamiltonian and any initial state, histories characterized by
projections onto any operators are decoherent if the widths of the
projections are taken to be sufficiently large, and the centers of
the projections are also suitably chosen. Furthermore, the
probabilities for these histories are then peaked about the time
evolution of the expectation value of the operators in the given
initial state.

At some level, this result must be almost trivial, since it is so
very general (``any'' initial state, ``any'' Hamiltonian, ``any''
operator). However, as we shall see, the hydrodynamic case
considered in this paper is a situation to which it naturally
applies, where it is little more than the statement that the quantum
evolution is strongly concentrated around one particular history.

To give a flavour of the basic idea, consider the simplest possible
case of a history characterized by a single projection at one moment
of time. Such histories are trivially decoherent, and the probabity
is given by the standard expression,
$$
p(\a ) = \Tr \left( P_{\a} \rho \right) = \la P_{\a} \ra 
\eqno(A1)
$$
For simplicity, we will let the projection $P_{\a}$ be a Gaussian
quasi-projector onto some operator $A$ with width $\s$:
$$
P_{\a} = {1 \over ( 2 \pi \s^2)^{\half} }
\ \exp \left( - {( A - \a )^2 \over 2 \s^2 } \right)
\eqno(A2)
$$
Gaussian projectors are exhaustive but only approximately exclusive.

Inserting this in (A1) and letting
$ \b = \a - \la A \ra $, and $B = A - \la A \ra $, we have
$$
p(\a ) = {1 \over ( 2 \pi \s^2)^{\half} }
\ \la \exp \left( - { ( \b - B )^2 \over 2 \s^2 } \right) \ra
\eqno(A3)
$$
We are interested in the behaviour of this expression for large
$\s$. Expanding for large $\s$ we obtain,
$$
p(\a) = {1 \over ( 2 \pi \s^2)^{\half} }
\ \exp \left( - { \b^2 \over 2 \s^2} \right)
\ \la 1 + { \b B \over \s^2 } - { B^2 \over 2 \s^2 }
+ { \b^2 B^2 \over 2 \s^4 } + O (\s^{-3} ) \ra
\eqno(A4)
$$
(note that $\b$ is taken to be of order $\s$ because of
the Gaussian factor). Now, noting that $ \la B \ra = 0$,
and $ \la B^2 \ra = ( \Delta A)^2 $, and reorganizing the result
as an exponential, we find that for large $\s$,
$$
p(\a) \ \approx 
\ { 1 \over ( 2 \pi ( \s^2 + (\Delta A)^2 ) )^{\half} }
\ \exp\left( - {( \la A \ra - \a_1 )^2 \over 2 (\s^2 
+ (\Delta A)^2 )}  \right)
\eqno(A5)
$$
Therefore, for $ \s^2 >> (\Delta A)^2 $, the probability 
is peaked about the expectation value of $A$. This is an intuitively
expected result: a very broad sampling yields the average value.
(The same qualitative result will probably also hold for exact
projectors, but we do not consider this here).

The above result readily extends to 
non-trivial histories. First note that for Gaussian
projectors, the time-dependent projectors may be written,
$$
P_{\a_k} (t_k) = {1 \over ( 2 \pi \s^2)^{\half} }
\ \exp \left( - {( A_{t_k} - \a_k )^2 \over 2 \s^2 } \right)
\eqno(A6)
$$
where  
$$
A_{t_k} = e^{i H(t_k-t_0)} A  e^{-i H(t_k-t_0)} 
\eqno(A7)
$$
Then inserting in the expression for the probabilities 
(setting aside for the moment the question of decoherence)
and expanding for large $\s$, the leading order behaviour is,
$$
p (\a_1, \a_2, \cdots \a_n ) \ =
\ \prod_{k=1}^n
\ \exp \left( - {( \la A_{t_k} \ra - \a_k )^2 \over \s^2 } \right)
\eqno(A8)
$$
For simplicity, we have ignored normalization factors. Also, we have
kept only the leading order terms for  $ \s^2 >> ( \Delta A_{t_k}
)^2 $ (for times $t_k$). The next correction may be computed, at
some length, but the qualitative conclusion is the same. 

Eq.(A8) indicates that the probabilites for histories are peaked
about the evolution of the expectation value, $ \la A_t \ra $.
It is the natural generalization of (A5): the broadest possible
sampling of a history, yields the evolution of the expectation value.

The result for the single time case
extends easily to the multi-time case even though the $A_{t_k}$'s
do not commute, because their non-commutativity only enters at
higher orders for large $\s$, and does not affect the configuration
$ \la A_{t_k} \ra $ about which the probabilities are peaked.
For the same reason, one can also project onto a number of 
different non-commuting operators at each moment of time, 
as one needs to in the hydrodynamic case (see (5.7)--(5.9)),
and still obtain a result of the form (A8).

The argument for decoherence is similar to those in Sections IV and
V -- it arises because there is essentially only one history with
non-zero probability. More precisely, define a set containing just
two histories as follows. Take the first history to be that 
characterized by projections in which the $\a_k$'s are chosen to be
very close to $ \la A_{t_k} \ra $, and the width $\s^2 $ is much
greater than $ ( \Delta A)^2 $. Then the probability for his
history, which we denote simply $p$, will be very close to $1$. Then
take the remaining history to be the complement of the first one.
That is, the combination of all histories defined by projections in
which at least one of the $\a_k$'s is much greater than the width
$\s$ away from $ \la A_{t_k} \ra $. Then the probabilities for the
second history, denote it $\bar p$, will be close to zero, since at
least one of the projections will always pick up an exponentially
small tail. There will therefore be approximate consistency, since
the probabilites approximately add up to $1$. (Equivalently, 
the decoherence functional is very small, since $ | D |^2 < p
\bar p $.)

As stated at the beginning, the above result is very general to the
point of being almost vacuous. But it is useful in the hydrodynamic
case because the local densities do have extremely small variances $
(\Delta A)^2 $ (compared to $ \la A \ra^2 $) in the local
equilibrium state, so the widths $\s$ do not need to be taken to be
unreasonably large for the conditions of the theorem to be
satisfied. Hence we have provided an alternative proof that the
probabilities for histories of local densities are peaked about the
hydrodynamic equations.

\references
\def\pr{{\sl Phys.Rev.}}

\refis{Ana} C.Anastopoulos,
``Hydrodynamic equations from quantum field theory",
preprint gr-qc/9805074 (1998).

\refis{AnH} C.Anastopoulos and J.J.Halliwell, 
{\sl Phys.Rev.} {\bf D51}, 6870 (1995).

\refis{BrH} T.Brun and J.J.Halliwell, 
{\sl Phys.Rev.} {\bf 54}, 2899 (1996).

\refis{BrHa} T.Brun and J.B.Hartle, ``Hydrodynamic Equations from
Decoherent Histories'', Santa Barbara preprint (in preparation).

\refis{CaH} E. Calzetta and B. L. Hu, in {\it Directions in General
Relativity}, edited by B. L. Hu and T. A. Jacobson (Cambridge
University Press, Cambridge, 1993).

\refis{CaL} A.Caldeira and A.Leggett, {\sl Physica} {\bf 121A}, 587 (1983).

\refis{Car} B.Carazza, ``On the spatial density matrix for the centre
of mass of a one-dimensional perfect gas'', preprint
quant-ph/9711005 (1997).

\refis{Dio} L.Di\'osi, preprint gr-qc/9403046.

\refis{DoH} H. F. Dowker and J. J. Halliwell, {\sl Phys. Rev.} {\bf
D46}, 1580 (1992).

\refis{DoK} H.F.Dowker and A.Kent, {\sl J.Stat.Phys.} {\bf 82},
1575 (1996); {\sl Phys.Rev.Lett.} {\bf 75}, 3038 (1995).

\refis{eqm} The approach to equilibrium (or local equilibrium) is
discussed in many places. Some examples are,
G.S.Agarwal, \pr {\bf A3}, 838 (1971);
M.A.Huerta and H.S.Robertson, {\sl J.Stat.Phys.} {\bf 1}, 393 (1969);
M.Tegmark and H.S.Shapiro, \pr {\bf E50}, 2538 (1994).

\refis{FeV} R.P.Feynman and F.L.Vernon, 
{\sl Ann. Phys.} {\bf 24}, 118 (1963).

\refis{Fef} C.L.Feffermann, {\sl Bull.Amer.Math.Soc} {\bf 9}, 129
(1983). 

\refis{FGH} D.Finkelstein, {\sl Trans.N.Y.Acad.Sci.} {\bf 25}, 
621 (1963);
N.Graham, in {\it The Many Worlds Interpretation of
Quantum Mechanics}, B.S.DeWitt and N.Graham (eds.) (Princeton
University Press, Princeton, 1973);
J.B.Hartle, {\sl Am.J.Phys.} {\bf 36}, 704 (1968).
See also, E.Farhi, J.Goldstone and S.Gutmann, 
{\sl Ann.Phys.(NY)} {\bf 192}, 
368 (1989).

\refis{Flo} J.C.Flores, ``Decoherence from internal degrees of
freedom for clusters of mesoparticles: a hierachry of master
equations'', preprint quant-ph/9803032 (1998).

\refis{GH1} M.Gell-Mann and J.B.Hartle, in {\it Complexity, Entropy 
and the Physics of Information, SFI Studies in the Sciences of Complexity},
Vol. VIII, W. Zurek (ed.) (Addison Wesley, Reading, 1990); and in
{\it Proceedings of the Third International Symposium on the Foundations of
Quantum Mechanics in the Light of New Technology}, S. Kobayashi, H. Ezawa,
Y. Murayama and S. Nomura (eds.) (Physical Society of Japan, Tokyo, 1990).

\refis{GH2} M.Gell-Mann and J.B.Hartle, {\sl Phys.Rev.} {\bf D47},
3345 (1993).


\refis{GiR} I.Giardina and A.Rimini, {\sl Found.Phys.} {\bf 26}, 
973 (1996).

\refis{Gri} R.B.Griffiths, {\sl J.Stat.Phys.} {\bf 36}, 219 (1984);
{\sl Phys.Rev.Lett.} {\bf 70}, 2201 (1993).

\refis{Gri2} R.B.Griffiths, {\sl Phys.Rev.} {\bf A54}, 2759 (1996);
{\bf A57}, 1604 (1998).

\refis{Hal1} J.J.Halliwell, in {\it Fundamental Problems in Quantum
Theory},  edited by D.Greenberger and A.Zeilinger, Annals of the New
York Academy of Sciences, Vol 775, 726 (1994).

\refis{HaZ} J.J.Halliwell and A.Zoupas, {\sl Phys.Rev.}
{\bf D52}, 7294 (1995); {\bf D55}, 4697 (1997).


\refis{Har6} J. B. Hartle, ``Quasiclassical Domains in a Quantum 
Universe'', preprint gr-qc/9404017 (1994).

\refis{Har7} J.B.Hartle, ``Quantum Pasts and the Utility of 
Histories", preprint gr-qc/9712001.

\refis{HLM} J. B. Hartle, R. Laflamme and D. Marolf, 
\pr {\bf D51}, 7007 (1995).

\refis{hydro} The standard derivation of the hydrodynamic equations from
their underlying microscopic origins is described in many places. An
interesting discussion of the general issues is by G.E.Uhlenbeck, in
the Appendix of, {\it Lectures in Applied Mathematics: Probability
and Related Topics in the Physics Sciences}, by M.Kac (Interscience,
NY, 1957). A useful basic discussion may be found in
K.Huang, {\it Statistical Mechanics}, 2nd edition (New York,
Chichester, Wiley, 1987).
See also, for example,
R.Balescu, {\it Statistical Dynamics: Matter out of Equilibrium},
(Imperial College Press, London, 1997);
D. Forster, {\it Hydrodynamic Fluctuations, Broken
Symmetry and Correlation Functions} (Benjamin, Reading, MA, 1975);
H.J.Kreuzer, {\it Nonequilibrium Thermodynamics and its
Statistical Foundations} (Clarendon Press, Oxford, 1981);
D.N.Zubarev, {\it Nonequlibrium Statistical
Thermodynamics} (Consultants Bureau, New York, 1974);
D.Zubarev, V.Morozov and G.R\"opke, {\it
Statistical Mechanics of Nonequilibrium Processes, Vol.1}
(Akademie Verlag, Berlin, 1996).

\refis{JoZ} E.Joos and H.D.Zeh, {\sl Z.Phys.} {\bf B59}, 223 (1985).

\refis{Omn} R. Omn\`es, {\sl J.Stat.Phys.} {\bf 53}, 893 (1988).
{\bf 53}, 933 (1988);
{\bf 53}, 957 (1988);
{\bf 57}, 357 (1989);
{\sl Ann.Phys.} {\bf 201}, 354 (1990);
{\sl Rev.Mod.Phys.} {\bf 64}, 339 (1992).

\refis{Omn2} R.Omn\`es, comment after talk by W.Zurek
in {\it Physical Origins of Time Asymmetry},
edited by  J.J.Halliwell, J.Perez-Mercader and W.Zurek (Cambridge
University Press, Cambridge, 1994), page 209.

\refis{Omn3} R.Omn\`es, \pr {\bf A56}, 3383 (1997).

\refis{Ris} H.Risken, {\it The Fokker-Planck Equations: Methods of
Solution and Applications}, Second Edition (Springer-Verlag, Berlin,1989).

\refis{Zub} D.N.Zubarev, {\it Nonequilibrium Statistical
Thermodynamics} (Consultants Bureau, New York, 1974);

\refis{Zur} See for example, J.P.Paz and W.H.Zurek, \pr {\bf D48},
2728 (1993); W.Zurek, in {\it Physical Origins of Time Asymmetry},
edited by  J.J.Halliwell, J.Perez-Mercader and W.Zurek (Cambridge
University Press, Cambridge, 1994); W.Zurek, ``Decoherence,
Einselection, and the Existential Interpretation
(The Rough Guide)'', preprint quant-ph/9805065.

\endreferences

\end